\documentclass[5p,sort&compress,number]{elsarticle}

\usepackage{lineno}

\usepackage{url}
\makeatletter
\def\url@leostyle{%
  \@ifundefined{selectfont}{\def\UrlFont{\sf}}{\def\UrlFont{\footnotesize\ttfamily}}}
\makeatother
\newcommand{\be}{\begin{equation}} 
\newcommand{\ee}{\end{equation}}
\newcommand{\bea}{\begin{eqnarray}} 
\newcommand{\eea}{\end{eqnarray}}

\def\etal{\textit{et al.}}
\def\NC{\mathrm{NC}}
\begin{document}

%------------------------------------------------------------
\title{Data Acquisition, Triggering, and Filtering at the Auger
  Engineering Radio Array}

\author[ru]{J.~L.~Kelley}
\ead{j.kelley@astro.ru.nl}

\author[pao]{for the Pierre Auger Collaboration\fnref{paauth}}

\address[ru]{Dept.~of Astrophysics / IMAPP, Radboud University
  Nijmegen, 6500GL Nijmegen, Netherlands}
\address[pao]{Observatorio Pierre Auger, Av. San Mart\'in Norte 304, 5613 Malarg\"ue,
  Argentina}

\fntext[paauth]{Full author list:\\\url{http://www.auger.org/archive/authors_2011_10.html}}

%------------------------------------------------------------

\begin{abstract}
The Auger Engineering Radio Array (AERA) is currently detecting cosmic rays
of energies at and above $10^{17}$ eV at the Pierre Auger Observatory, by
triggering on the radio emission produced in the associated air showers.
The radio-detection technique must cope with a significant background of
man-made radio-frequency interference, but can provide information on
shower development with a high duty cycle.  We discuss our techniques to
handle the challenges of self-triggered radio detection in a low-power autonomous array, including
triggering and filtering algorithms, data acquisition design, and
communication systems.
\end{abstract}


%------------------------------------------------------------
\maketitle

%
% Line numbers for reviewing
%
%\modulolinenumbers[5]
%\setlength\linenumbersep{3pt}
%\linenumbers

%------------------------------------------------------------

\section{Introduction}

Cosmic-ray air showers emit coherent broadband radio pulses in the MHz
frequency range, a phenomenon discovered by Jelley and collaborators in
1965 \cite{jelley_radio}.  Advances in digital processing technology, along
with the potential of the radio technique to determine cosmic-ray primary
composition with a high duty cycle \cite{composition}, have led to renewed
interest in the technique \cite{falcke_nature}.  Radio enhancements are now
planned at several neutrino and cosmic-ray observatories.  

The goals of the Auger Engineering Radio Array (AERA), a 20-$\mathrm{km}^2$
extension of the Pierre Auger Observatory in Argentina, are to calibrate the
radio signals using hybrid air-shower measurements, and to demonstrate the
potential of the technique for future arrays.  The first
stage of AERA is operational and is successfully measuring air showers of
energies at and above $10^{17}$ eV in coincidence with the Auger surface
detector (SD) and fluorescence detector (FD) \cite{aera_icrc11}.  

\section{Detector and Data Acquisition Design}

The first stage of AERA, deployed in September 2010 and July 2011, consists
of 24 radio detector stations (RDSs) deployed on a triangular grid
with a spacing of 150 m and covering an area of $\sim0.4\ \mathrm{km}^2$.  Each RDS
measures the radio signal with a dual-polarization log-periodic dipole
antenna (LPDA), aligned north-south and east-west, and sensitive between 27
and 84 MHz.  The received signal is further amplified and bandpass
filtered, and each polarization is digitized at 200 MHz with 12-bit
analog-to-digital converters (ADCs).  High-gain (+49 dB after the
antenna) and low-gain (+29 dB) versions of each channel are recorded to
extend the dynamic range.

A field-programmable gate array (FPGA) in the station electronics forms a
trigger when a bandwidth-limited pulse is observed in the voltage
traces (a ``level 1'' trigger).
These triggers are passed to a CPU within the RDS, which can then
optionally apply other trigger conditions (the ``level 2'' trigger).  The
triggers are timestamped using GPS receivers at each RDS.

\begin{figure}
\centering
\includegraphics[scale=0.43,angle=-90]{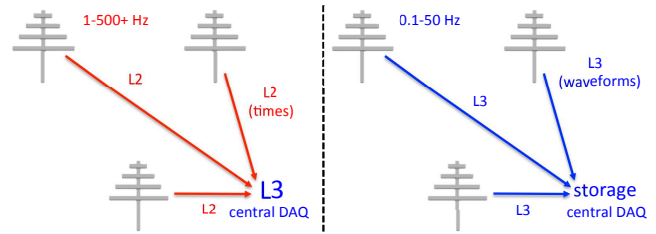}
\caption{The AERA data flow.  The GPS timestamps of station-level
  triggers (L2) are sent to a central DAQ, which forms multi-station
  coincidences (L3; shown at left).  An L3 request is sent back to the
  stations (not shown), and digitized waveforms of the event are sent to
  the central DAQ for storage (shown at right).}
\label{fig:daq}
\end{figure}

The CPU in each RDS communicates via Ethernet over a fiber-optic network to
a central data acquisition system (DAQ).  The RDS reports
the trigger timestamps to the central DAQ.  The DAQ software forms
the ``level 3'' trigger by using time and spatial coincidences of multiple
RDSs, and the time-domain waveform data of
these events are requested from each RDS and stored to disk.  Figure
\ref{fig:daq} shows the data flow for each trigger level.

\section{Triggering and Filtering}

Triggering directly on the radio signal of the air showers (instead of using
particle detectors as a trigger) poses some challenges for the data
acquisition, due to man-made radio-frequency interference (RFI).  The
continuous background level is set by the radio emission from the Galactic
plane, but any man-made narrowband transmitters add to the level above
which one must detect air-shower pulses.  Additionally, man-made pulsed RFI
(from sparking electrical equipment, airplanes, etc.) can mimic the signals
from cosmic rays.  

\begin{figure}
\begin{minipage}{0.98\linewidth}
\centering
\includegraphics[scale=0.45]{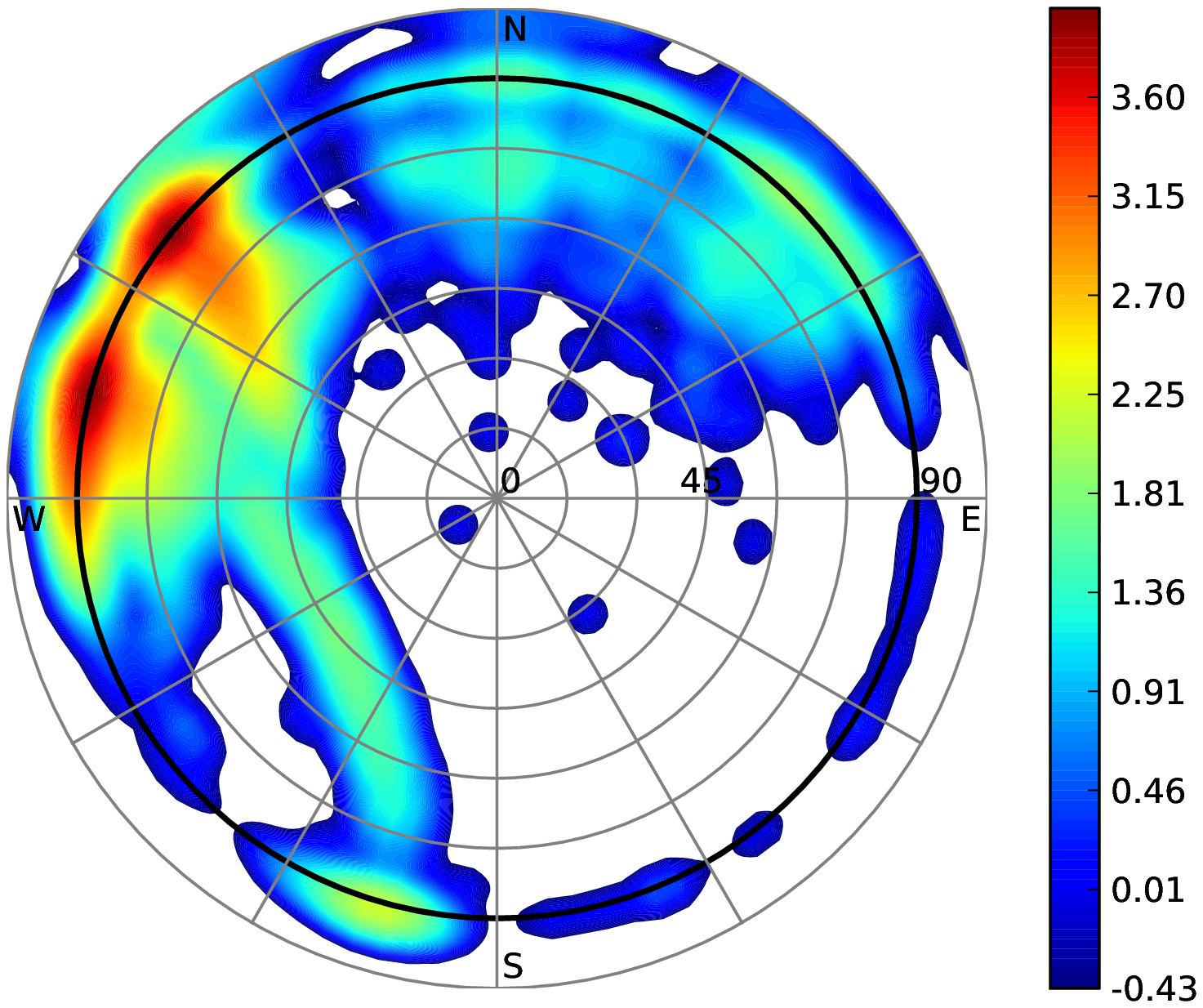}
\end{minipage}
\vskip-1.3em
\begin{minipage}{0.98\linewidth}
\centering
\includegraphics[scale=0.45]{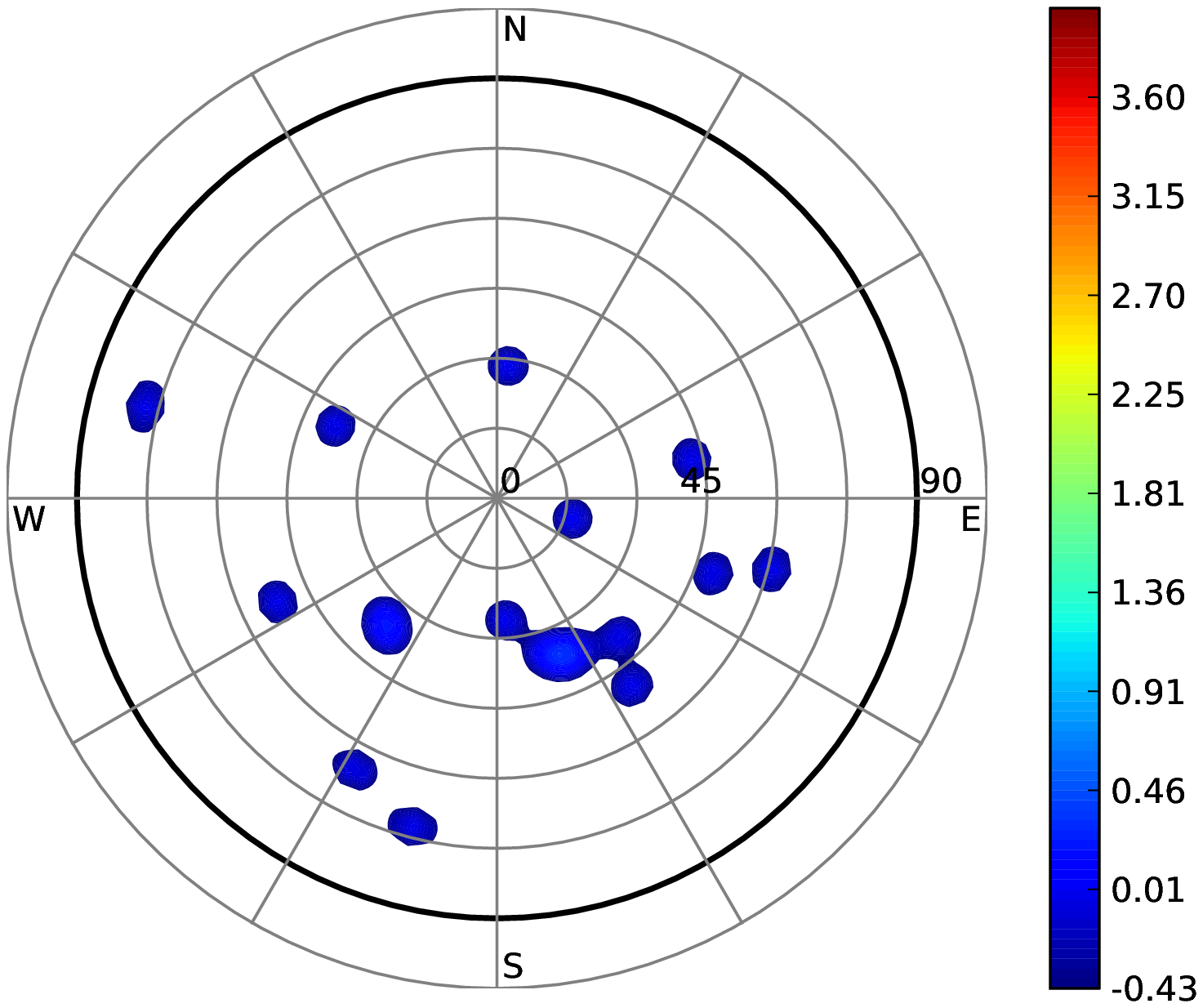}
\end{minipage}
\vskip-2em
\caption{Top: unfiltered AERA polar skymap (2.5h, 40k events).  The dark
  circle indicates the horizon.  Bottom: AERA
  cosmic-ray events coincident with the Auger SD (48d, 18 events).  The
  directions have been smeared with a $3^\circ$ Gaussian, and the color
  scale is $\log_{10}(\mathrm{event\ density, a.u.})$.} 
\label{fig:skymaps}
\end{figure}

A comparison is shown in Fig.~\ref{fig:skymaps} between raw level-3-triggered data and
self-triggered cosmic-ray events detected in coincidence with the SD.
Since the bandwidth and computational resources at each triggering level are
limited, one of the technical focuses for the first stage of the array has been to develop
various methods to reject RFI in order to minimize efficiency losses from
bandwidth saturation.   

\subsection{\label{sect:iir}Narrowband Filtering}

Before triggering on a radio pulse, it is advantageous to increase the
signal-to-noise ratio by filtering out any narrowband transmitters from the
digitized antenna signals.  This is accomplished in a computationally
efficient manner by using a series of infinite-impulse response (IIR) notch
filters in the FPGA.  The IIR filters operate on the time-domain signal,
and the output of the filter $y_i$ is a linear combination of input samples $x_j$ and
delayed feedback output samples $y_j$ from the filter:
\bea
y_i &=& x_i - (2\cos\omega_N \cdot x_{i-1}) + x_{i-2} \nonumber \\
&& + (2 r \cos\omega_N \cdot y_{i-1}) - (r^2 \cdot y_{i-2}) .
\label{eq:iir}
\eea

\noindent The normalized filter frequency $\omega_N$ is given by the notch frequency $f_N$ and the
sampling frequency $f_S$,

\be
\omega_N = 2 \pi f_N/f_S\ ,
\ee

\noindent and the width parameter $r$ is a value strictly between 0 and 1, with
higher values giving a narrower response function. $r=0.99$ is typical for
a narrow transmitter.

A complication in the implementation of the IIR filters in high-frequency
FPGAs arises from the fact that one cannot arbitrarily pipeline the
feedback computation.  We have resolved this by using the scattered
look-ahead pipelining technique \cite{parhi_scatter}, which increases the
filter complexity but allows more time for computation in the FPGA.  

\begin{figure}
\centering
\includegraphics[scale=0.4]{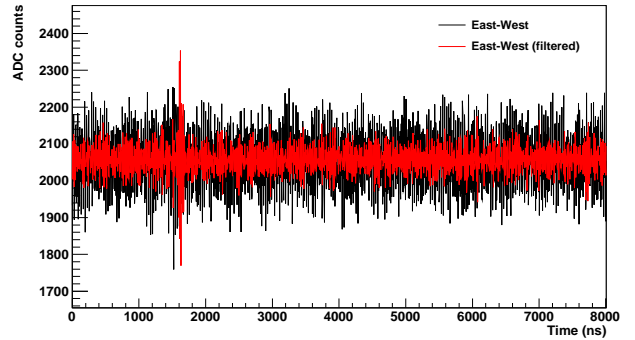}
\caption{A recorded cosmic-ray radio pulse (at 1500 ns), before IIR narrowband filtering
  (black) and after (red, pulse offset in time).  The peak signal-to-noise ratio
  increases from 2.4 to 7.6.}
\label{fig:notch}
\end{figure}

The coefficients of the
$x_j$ and $y_j$ in Eq.~\ref{eq:iir} can be precomputed for the desired
notch frequencies, converted to a fixed-point representation, and loaded
into the FPGA at run-time.  The current design allows for four tunable notch
filters for each polarization.  See Fig.~\ref{fig:notch} for an example of
the filtering.

An additional filtering technique we are exploring is the use of a
frequency-domain median filter.  The time-domain signal is first
converted to the frequency domain with a fast-Fourier transform (FFT), and then
the amplitude of each frequency component $f_i$ is replaced by the median
value in a window of size $2N$: $\tilde{f}_i = \mathrm{median}(f_{i-N},
..., f_{i+N})$.  An inverse Fourier transform then converts the signal back to the
time domain for triggering.  The median filter has the advantage of
removing any number of narrowband 
transmitters, but it is computationally intensive and can distort pulse
shapes in the time domain.  The pre-trigger FFT also allows for
a number of other strategies for signal-to-noise improvement, such as
deconvolution of the bandpass filter dispersion, or use of a matched filter
\cite{ruehle_arena2010}.  

\subsection{Level 1 Trigger}

Designed to capture transient, isolated radio pulses, the digitizer uses
algorithms in the FPGA to trigger in the time domain on potential signals.
At its most basic, the trigger is simply a voltage threshold above a
baseline.  However, a number of other parameters are added to this trigger
to reject RFI.  The following trigger conditions are
applied to select clean, bandwidth-limited pulses:

\begin{enumerate}
\item{the voltage rising edge crosses the primary threshold $T_1$;}

\item{before the $T_1$ crossing, no other $T_1$ crossings occur during the
  previous time period $T_{\mathrm{prev}}$;}

\item{after the $T_1$ crossing, the signal rising edge crosses a secondary
  threshold $T_2$, where normally $T_2 < T_1$;}

\item{the number of rising-edge $T_2$ crossings $\NC$ within a time period
  $T_{\mathrm{per}}$ falls within a specified
  range $\NC_{\mathrm{min}} < \NC < \NC_{\mathrm{max}}$;}

\item{the time $\mathrm{TC}$ between successive $T_2$ crossings is
  less than a maximum value $\mathrm{TC}_{\mathrm{max}}$;}

\item{the quotient $Q$ of the pulse maximum $P_{\mathrm{max}}$ divided by the
  number of $T_2$ crossings $\NC$ falls within the range $Q_\mathrm{min} < Q
  < Q_\mathrm{max}$\ .}
\end{enumerate}

\noindent The thresholds are compared relative to a baseline voltage,
determined dynamically with a rolling average over a 
specified number of samples.  Samples outside of a specified voltage range
(i.e.~part of a large pulse) are not included in the baseline calculation.  

\begin{figure}
\centering
\includegraphics[scale=0.7,angle=-90]{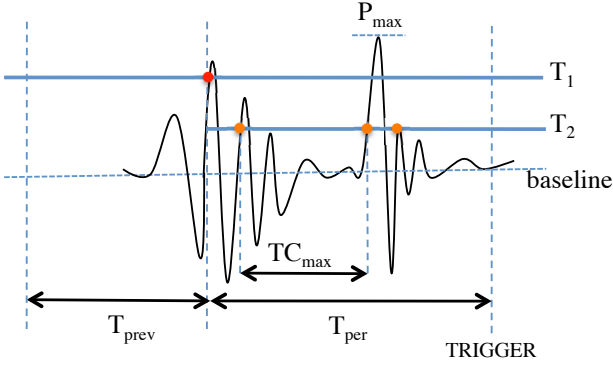}
\caption{The AERA L1 trigger scheme.  See the text for details.} 
\label{fig:trigger}
\end{figure}

Figure~\ref{fig:trigger} shows the trigger parameters graphically.  Each
condition is designed to reject man-made RFI with specific
characteristics.  The quiet period $T_{\mathrm{prev}}$ rejects pulse
trains.  The secondary threshold $T_2$, which is generally set lower than
the primary threshold $T_1$, rejects signals with after-pulsing, even if the
after-pulsing is weaker than the initial pulse.  The condition on the timing
and number of $T_2$ crossings allows for some ringing of a bandwidth-limited pulse
but rejects long and/or irregular pulse trains or digital spikes.  The
condition on $Q = P_{\mathrm{max}} / \NC$ can select on 
pulse shape, but in practice this is not needed when the other trigger
parameters are set correctly.   

Typical values for the thresholds and noise rejection parameters are as
follows: $T_1 = 200$ ADC counts and $T_2 = 150$ ADC counts, on an RMS noise
level of 40 ADC counts; $\NC_{\mathrm{min}} = 1$ and $\NC_{\mathrm{max}} =
8$ during a period of $T_{\mathrm{per}} = 6.25\ \mu\mathrm{s}$; maximum
crossing time $\mathrm{TC}_{\mathrm{max}}$ of 130 ns; and a quiet
pre-trigger period of $T_{\mathrm{prev}} = 1.25\ \mu\mathrm{s}$. 

The algorithm can trigger on either signal polarization, or the logical AND
or OR of the two.  The narrowband filtering described in Sect.~\ref{sect:iir} is
performed before the trigger algorithm, but one can also
trigger on the unfiltered signal; the unfiltered signal is stored
in either case.  

\subsection{Periodicity Filtering}

Man-made pulsed RFI, unlike cosmic-ray air showers, is often periodic in
nature.  In particular, by examining the time between successive level 1
triggers, clear signatures are visible at 100 Hz and sub-harmonics thereof
(twice the power grid frequency of Argentina).   

\begin{figure}
\centering
\includegraphics[scale=0.30]{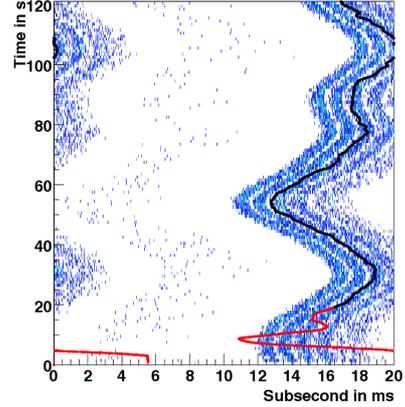}
\caption{Sample L1 triggers, showing the residual from a 20 ms period
  (50 Hz) vs.~time.  The drifting phase can be tracked with a digital PLL
  (black line).}
\label{fig:50hz}
\end{figure}

One difficulty in vetoing such events is the drift in phase of the 100 Hz
pulses.  This can be tracked in the RDS software with the use of a digital
phase-locked loop, which can track the changing phase and allow a veto of the
periodic events (see Fig.~\ref{fig:50hz}).

\subsection{Directional Filtering}

As shown in Fig.~\ref{fig:skymaps}, most of the pulsed background events
come from man-made sources on the horizon.  When a level 3 trigger
is formed with three or more stations in coincidence, the DAQ has enough information to
reconstruct the direction and veto these events, either with a simple
selection on the reconstructed zenith angle, or by using additional azimuthal
and timing information.  The only limitation is the computation time required
for a directional reconstruction, which must keep up with the event rate.
A plane-wave fit to the trigger times of the stations is sufficiently fast
and accurate for such purposes.

\begin{figure}
\centering
\includegraphics[scale=0.35]{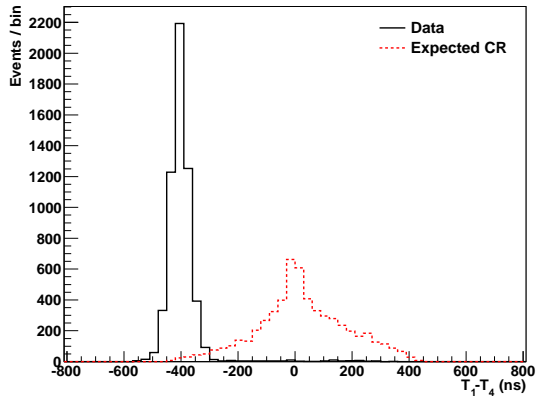}
\caption{Trigger-time difference between stations 1 and 4 (solid:
  sample data, showing a pulsed RFI point source; dashed: expected
  cosmic-ray distribution, normalized to the same number of events).} 
\label{fig:delta_t}
\end{figure}

A faster method that can veto directional hot spots, but does not require
a full directional reconstruction, uses the distribution of trigger time
differences between two stations.  Directional hot spots can be identified
by peaks in the time difference histograms for each station pair, and
these distributions are quite different from that expected from air showers
(see Fig.~\ref{fig:delta_t}).  

To remove the hot spots, the DAQ forms dynamic
histograms $F(i,j)$ of trigger time differences $T_i - T_j$ for each station pair
$(i,j)$; the histograms continuously update using a fixed number of past entries.  For a
candidate level 3 trigger event, the number of ``good'' time differences $(T_i, T_j)$
is calculated as 

\be
N_{\mathrm{good}}\ =\sum_{N_{\mathrm{pairs}}} \left\{ 
\begin{array}{ll}
1 & \mathrm{if}\ F(T_i-T_j) < F_c \\
0 & \mathrm{otherwise}
\end{array}
\right.\ ,
\ee

\noindent where $F_c$ is an adjustable threshold that depends on the bin size
and number of entries in the histograms.  The selection criterion to keep
the event is $N_{\mathrm{good}} > N_{\mathrm{pairs}} / 2\ $.  Future
revisions may include the expected time difference distribution of
air-shower events as well.   

\subsection{Offline Filtering}

To date, confirmed cosmic-ray events in AERA have been found by searching
offline for time coincidences with the water-Cherenkov surface detector of
the Pierre Auger Observatory.  However, we are developing methods to
identify cosmic-ray candidates without information from other detectors.

One promising technique is the use of the signal
polarization.  Since the primary component of the air-shower radio emission
is geomagnetic in nature \cite{polarization}, the
direction of the predicted electric field $\hat{e}_g$ is parallel to $\hat{v}
\times \hat{b}$, where $\hat{v}$ is the direction of the shower axis, and
$\hat{b}$ is the direction of the local magnetic field of the Earth.  By
deconvolving the response of the antenna and the station electronics, the
measured electric field of a radio pulse $\vec{E}(t)$ can be reconstructed
\cite{radio_offline}.  The direction of polarization can be compared to
that expected from an air shower with purely geomagnetic emission, i.e.,
events are selected if

\be
\cos^{-1} \left( \hat{e}_g \cdot
\frac{\vec{E}_{\mathrm{max}}}{|\vec{E}_{\mathrm{max}}|}\ \right) < \Psi_c\ ,
\ee

\noindent where $\Psi_c$ is a threshold space angle.  This method does not
yet account for the contribution of secondary emission mechanisms with
different polarization signatures, such as that
from the charge excess in the shower \cite{schoorlemmer_arena2010,
  codalema_icrc2011}; however, this can in principle be added using
additional information such as the core position of the shower.

\section{Future Plans}

AERA will expand in 2012 and 2013 to instrument an area of 20
$\mathrm{km}^2$ with 160 radio detector stations.  This expansion will
entail a transition from communication over optical fiber to a wireless
network.  Several designs are under development for the wireless system,
including:

\begin{enumerate}
\item{a fully custom time division multiple access (TDMA) system in the
  2.4 GHz band that supports up to 180 subscribers per channel, at a bandwidth of 5.5 Mbps;}
\item{a commercial 802.11n + TDMA system in the 5 GHz band that supports
  80-100 subscribers per channel, at a peak bandwidth of 150 Mbps (80 Mbps
  typical TCP/IP throughput); and}
\item{a distributed ``gossip'' protocol in which stations communicate with
  each other and form multi-station coincidences without communication with
  a central DAQ.}
\end{enumerate}

\noindent The first two systems are currently being field tested at the
AERA site, while the third system is in the development and
simulation stage.  

In addition to the developments in data acquisition, triggering, filtering,
and communications systems described here, the future stages of AERA will
also use enhanced antennas, low-noise amplifiers, and digitization
hardware.  The increase in efficiency as well as detection area will result
in a significant sample of hybrid and super-hybrid events that will be
used to quantify the potential of the radio technique for measurement of
cosmic-ray air shower energy, direction, and primary composition.

%------------------------------------------------------------

\clearpage
\end{document}